\documentclass[aps,preprint]{revtex4}%
\usepackage{amsfonts}
\usepackage{amsmath}
\usepackage{amssymb}%
\begin{document}
\preprint{CTP-SCU/2019007}
\title{ Thermodynamics and weak cosmic censorship conjecture in the charged RN-AdS
black hole surrounded by quintessence under the scalar field }
\author{Wei Hong$^{a}$}
\email{thphysics_weihong@stu.scu.edu.cn}
\author{Benrong Mu$^{b}$}
\email{benrongmu@cdutcm.edu.cn}
\author{Jun Tao$^{a}$}
\email{taojun@scu.edu.cn}
\affiliation{$^{a}$Center for Theoretical Physics, College of Physics, Sichuan University,
Chengdu, 610064, China}
\affiliation{$^{b}$Physics Teaching and Research Section, College of Medical Technology,
Chengdu University of Traditional Chinese Medicine, Chengdu 611137, China}

\begin{abstract}
In this paper, we study the thermodynamics and the weak cosmic censorship
conjecture of the RN-AdS black hole surrounded by the quintessence under the
scattering of a charged complex scalar field. With scalar field scattering,
the variation of the black hole is calculated in the extended and normal phase
spaces. In the extended phase space, the cosmological constant and the
normalization parameter are considered as thermodynamic variables, and the
first law of thermodynamics is valid, but the second law of thermodynamics is
not valid. In the normal phase space, the cosmological constant and the
normalization parameter are fixed, and the first and second laws of
thermodynamics can also be satisfied. Furthermore, the weak cosmic censorship
conjecture is both valid in the extended and normal phase spaces.

\end{abstract}
\maketitle
\tableofcontents

\section{Introduction}

Since Hawking and Bekenstein
\cite{Hawking:1974sw,Bekenstein:1974ax,Bekenstein:1973ur} proposed the black
hole thermodynamics, the research on the thermodynamics of black holes has
developed rapidly. Analogous to the laws of thermodynamics, the four laws of
black hole thermodynamics were established in \cite{Bardeen}. There are
several ways to calculate the temperature and entropy of black holes
\cite{tHooft:1984kcu,Parikh:1999mf,Hartle:1976tp}. A general argument gave
that the Hawking temperature of a black hole is proportional to the surface
gravity at the horizon, and the black hole entropy is proportional to the area
of the horizon. The black hole's horizon is an important part for the study of
the nature of black holes. According to Penrose's theory \cite{Penrose:1969pc}%
, all singularities caused by gravitational collapse must be hidden in the
black hole. In other words, singularities need to be hidden from an observer
at infinity by the event horizon of black hole, which is the weak cosmic
censorship conjecture.

The thermodynamic laws and the weak cosmic censorship conjecture can be tested
by the scattering of an external field or by the absorptions of a particle
through a black hole. As the particles fall into the black hole, it has been
proven that the first law and the second law of black hole thermodynamics are
still established \cite{Gwak:2016gwj,Gwak:2015fsa}. On the other hand, there
are some controversies about the weak cosmic censorship conjecture. Wald
innovatively proposed a method to test this conjecture in the extreme
Kerr-Newman black hole by absorbing a particle, which showed that the
conjecture was satisfied \cite{Wild:1974gx}. However, this conjecture is
violated in the near-extreme Reissner-Nordstr\"{o}m black hole
\cite{Hubeny:1998ga} and the near-extreme Kerr black hole
\cite{Jacobson:2009kt}. The validity of weak cosmic censorship conjecture in
the context of various black holes via the absorption of a charged or rotating
particle has been discussed by many work
\cite{Rocha:2011wp,Rocha:2014jma,Colleoni:2015afa,BouhmadiLopez:2010vc,Gao:2012ca,Gwak:2017icn,Duztas:2016xfg,Hod:2016hqx,Natario:2016bay,Horowitz:2016ezu,Gwak,Chen01,Chen02,Gwak:2017kkt,Wang:2019dzl,Kastor:2009wy,Chen:2018yah,Zeng:2019jrh,Zeng:2019jta}. Although a lot of work has been done, no consistent conclusion has been reached.

In this paper, we study the thermodynamics and weak cosmic censorship
conjecture of the RN-AdS black hole surrounded by the quintessence under the
scattering of a charged complex scalar field. The first law of thermodynamics
is always satisfied, but the second law of thermodynamics is not always
satisfied. And the weak cosmic censorship conjecture does not violate. The
paper is organized as follows. In section II, we investigate the dynamical of
the charged complex scalar field, and calculated the variations of this black
hole's energy and charge within the certain time interval. In section III, the
thermodynamics of the black hole are discussed in the extended and normal
phase space. Moreover, we test the validity of the weak cosmic censorship
conjecture in those case. We summarize our results in section IV.

\section{Complex scalar field in RN-AdS black hole surrounded by quintessence}

In this section, we first briefly review the RN-AdS black hole solution with
quintessence and then investigate the complex scalar field in the black hole background.

\subsection{Black hole solution}

The bulk action for a RN-AdS black hole surrounded by quintessence dark energy
in four dimensional curved space time could be described as follows
\cite{kvv2003, Ghaffarnejad:2018bsd}
\begin{equation}
\mathcal{S}=\frac{1}{16\pi G}\int d^{4}x(\sqrt{-g}\left[  R-2\Lambda-F^{\mu
\nu}F_{\mu\nu}\right]  +\mathcal{L}_{q}).
\end{equation}
In the above action, the cosmological constant is related to the AdS space
radius $l$ by $\Lambda=-3/l^{2}$. The last term $\mathcal{L}_{q}$ in the
action is the Lagrangian of quintessence as a barotropic perfect fluid, which
is given by \cite{Olivier}
\begin{equation}
\mathcal{L}_{q}=-\rho_{q}\left[  1+\omega_{q}\ln\left(  \frac{\rho_{q}}%
{\rho_{0}}\right)  \right]  ,
\end{equation}
where $\rho_{q}$ is energy density, $\rho_{0}$ is the constant of integral,
and the barotropic index $\omega_{q}$. One has that $-1<\omega_{q}<-1/3$ for
the quintessence dark energy and $\omega_{q}<-1$ for the phantom dark energy.
The metric of the charged RN-AdS black hole surrounded by quintessence is
\begin{align}
ds^{2} &  =f(r)dt^{2}-\frac{1}{f(r)}dr^{2}-r^{2}(r)\left(  d\theta^{2}%
+\sin^{2}\theta d\varphi^{2}\right)  ,\nonumber\\
f(r) &  =1-\frac{2M}{r}+\frac{Q^{2}}{r^{2}}-\frac{a}{r^{3\omega_{q}+1}}%
+\frac{r^{2}}{l^{2}}\text{,}\label{e2}%
\end{align}
where $M$ and $Q$ are the mass and electric charge of the black hole,
respectively, and $a$ is the normalization factor related to the density of
quintessence as \cite{AzregAinou:2012hy,Azreg-Ainou:2014lua}
\begin{equation}
\rho_{q}=-\frac{a}{2}\frac{3\omega_{q}}{r^{3}\left(  \omega_{q}+1\right)  }.
\end{equation}
The electromagnetic potential of the black hole is
\begin{equation}
A_{\mu}\left(  r\right)  =(-\frac{Q}{r},\ 0,\ 0,\ 0).\label{e3}%
\end{equation}
The potential of the black hole is
\begin{equation}
\varphi=-A_{t}\left(  r_{+}\right)  =\frac{Q}{r_{+}}.\label{varphi}%
\end{equation}
Moreover, the Hawking temperature can be derived as
\begin{equation}
T=\frac{f^{\prime}(r_{+})}{4\pi}=\frac{1}{4\pi}\left[  \left(  3\omega
_{q}+1\right)  r_{+}^{-3\omega_{q}-2}+\frac{2M}{r_{+}^{2}}+\frac{2r_{+}}%
{l^{2}}-\frac{2Q^{2}}{r_{+}^{3}}\right]  ,\label{fT}%
\end{equation}
and the entropy of the black hole is
\begin{equation}
S=\pi r_{+}^{2}.\label{e11}%
\end{equation}
In the extended phase space where the cosmological constant is treated as a
thermodynamic variable, the first law of thermodynamics via the RN-AdS black
hole surrounded by quintessence have been studied in \cite{Li:2014ixn}. The
mass of the black hole is then interpreted as the enthalpy
\begin{equation}
M=U+PV,\label{a}%
\end{equation}
where $U$ is internal energy.

\subsection{Complex scalar field}

The action of a complex scalar field in the fixed RN-AdS gravitational and
electromagnetic fields is
\begin{equation}
\mathcal{S}=-\frac{1}{2}\int\sqrt{-g}\left[  \left(  \partial^{\mu}-iqA^{\mu
}\right)  \Psi^{\ast}\left(  \partial_{\mu}+iqA_{\mu}\right)  \Psi-m^{2}%
\Psi^{\ast}\Psi\right]  d^{4}x,\label{eAction}%
\end{equation}
where $A_{\mu}$ is the electromagnetic potential, $m$ is the mass, $q$ is the
charge, $\Psi$ denotes the wave function, and its conjugate is $\Psi^{\ast}$.
Now we turn to investigate the dynamical of the charged complex scalar field.
The field equation obtained from the action satisfies
\begin{equation}
\left(  \nabla^{\mu}-iqA^{\mu}\right)  \left(  \nabla_{\mu}-iqA_{\mu}\right)
\Psi-m^{2}\Psi=0.\label{e18}%
\end{equation}
To solve this wave function, we carry out a separation of variables
\begin{equation}
\Psi=e^{-i\omega t}R(r)\Phi(\theta,\phi).\label{e19}%
\end{equation}
In the above equation, $\omega$ is the energy of the particle, and
$\Phi(\theta,\phi)$ is the scalar spherical harmonics. We put Eq. $\left(
\ref{e19}\right)  $ into Eq. $\left(  \ref{e18}\right)  $ and obtain the
radial wave function
\begin{equation}
R(r)=e^{\pm i(\omega-qQ/r)r_{\ast}},\label{e24}%
\end{equation}
where $dr_{\ast}=\frac{1}{f}dr$, $r_{\ast}$ is a function of $r$, and $+/-$
corresponds to the solution of the outgoing/ingoing radial wave. Since the
thermodynamics and the validity of the weak cosmic censorship conjecture are
discussed by the scattering of the ingoing wave at the event horizon in this
paper, we focus our attention on the ingoing wave function.

From the action $\left(  \ref{eAction}\right)  $, the energy-momentum tensor
is obtained as follows
\begin{equation}
T_{\nu}^{\mu}=\frac{1}{2}[(\partial^{\mu}-iqA^{\mu})\Psi^{\ast}\partial_{\nu
}\Psi+(\partial^{\mu}+iqA^{\mu})\Psi\partial_{\nu}\Psi^{\ast}]+\delta_{\nu
}^{\mu}\mathcal{L}.\label{e26}%
\end{equation}
Combining the ingoing wave function and its conjugate with the energy-momentum
tensor yields the energy flux
\begin{equation}
\frac{dE}{dt}=\int{T_{t}^{r}\sqrt{-g}d\theta d\phi}=\omega(\omega
-q\varphi)r_{+}^{2}.\label{e27}%
\end{equation}
The electric current is obtained from the action $\left(  \ref{eAction}
\right)  $
\begin{equation}
j^{\mu}=\frac{\partial\mathcal{L}}{\partial A_{\mu}}=-\frac{1}{2}iq[\Psi
^{\ast}(\partial^{\mu}+iqA^{\mu})\Psi-\Psi(\partial^{\mu}-iqA^{\mu})\Psi
^{\ast}].\label{e28}%
\end{equation}
For the ingoing wave function $\left(  \ref{e24}\right)  $, the charge flux
is
\begin{equation}
\frac{dQ}{dt}=-\int{j^{r}\sqrt{-g}d\theta d\varphi}=q(\omega-q\varphi
)r_{+}^{2}.\label{e29}%
\end{equation}

When the complex scalar field is scattered off the black hole, the decreases
of the energy and charge of the scalar field are equal to the increases of
these of the black hole due to the energy and charge conservation. From Eqs.
$\left(  \ref{e27}\right)  $ and $\left(  \ref{e29}\right)  $, the transferred
energy and charge within the certain time interval are
\begin{equation}
dU=dE=\omega(\omega-q\varphi)r_{+}^{2}dt,\ \ dQ=q(\omega-q\varphi)r_{+}%
^{2}dt,\label{e30}%
\end{equation}
respectively. Since the transferred energy and charge are very small, the time
$dt$ must be also very small. The increase or decrease of $dU$ and $dQ$ depend
on the relation between $\omega$ and $q\varphi$.

\section{Thermodynamics and weak cosmic censorship conjecture}
In this section, we test the thermodynamics laws and the weak cosmic
censorship conjecture of the black hole by the scattering of the ingoing wave
at the event horizon.

\subsection{Extended phase space}
In the extended phase space, the cosmological constant and normalization
parameters are considered as thermodynamic variables. One can treat the
cosmological constant as thermodynamic pressure and its conjugate quantity as
thermodynamic
volume.\cite{Dolan:2011xt,Cvetic:2010jb,Kubiznak:2012wp,Hendi:2012um,Azreg-Ainou:2014twa,Caceres:2015vsa,Pedraza:2018eey} The
definitions are as follows
\begin{equation}
P=-\frac{\Lambda}{8\pi}=\frac{3}{8\pi l^{2}},\\
V=\left(  \frac{\partial M}{\partial P}\right)  _{S,Q},\label{e6}%
\end{equation}
where $M$ is the mass of the black hole. So, the metric function $f(r)$ can be
rewritten as
\begin{equation}
f(r)=1-\frac{2M}{r}+\frac{Q^{2}}{r^{2}}-\frac{a}{r^{3\omega_{q}+1}}+\frac
{8}{3}\pi Pr^{2}.\label{e7}%
\end{equation}
Solving the equation $f(r)=0$ at the horizon radius $r=r_{+}$, one can obtain
the mass of the black hole
\begin{equation}
M=\frac{1}{6}\left(  -3ar_{+}^{-3\omega_{q}}+8\pi Pr_{+}^{3}+\frac{3Q^{2}
}{r_{+}}+3r_{+}\right)  .\label{e8}%
\end{equation}
Using the Eqs. $\left(  \ref{e6}\right)  $ and $\left(  \ref{e8}\right)  $, we
can obtain the thermodynamic volume as
\begin{equation}
V=\frac{4\pi r_{+}^{3}}{3}.\label{e9}%
\end{equation}

The initial state of the black hole is represented by $(M,Q,r_{+},P,a)$, and
the final state is represented by $(M+dM,Q+dQ,r_{+}+dr_{+},P+dP,a+da)$. The
variation of the horizon radius can be obtained from the variation of metric
function $f\left(  r_{+}\right)  $. For the initial state $(M,Q,r_{+},P,a)$
satisfies
\begin{equation}
f(M,Q,r_{+},P,a)=0.\label{fn}%
\end{equation}
When the black hole mass and charge are varied, we assume that the final state
of the charged RN-AdS black hole surrounded by quintessence is still a black
hole, which satisfies
\begin{equation}
f(M+dM,Q+dQ,r_{+}+dr_{+},P+dP,a+da)=0.\label{fnd}%
\end{equation}

The functions $f(M+dM,Q+dQ,r_{+}+dr_{+},P+dP,a+da)$ and $f(M,Q,r_{+},P,a)$
satisfy the following relation
\begin{align}
& f\left( M+d M, Q+d Q, r_{+}+d r_{+}, P+d P, a+d a\right) =f\left( M, Q,
r_{+}, P, a\right) \nonumber\\
& +\left. \frac{\partial f}{\partial M}\right| _{r=r_{+}} d M+\left.
\frac{\partial f}{\partial Q}\right| _{r=r_{+}} d Q+\left. \frac{\partial
f}{\partial r_{+}}\right| _{r=r_{+}} d r_{+}+\left. \frac{\partial f}{\partial
P}\right| _{r=r_{+}} d P+\left. \frac{\partial f}{\partial a}\right|
_{r=r_{+}} d a,\label{fextend}%
\end{align}
where
\begin{equation}
\left. \frac{\partial f}{\partial M}\right| _{r=r_{+}}=-\frac{2}{r_{+}},\left.
\frac{\partial f}{\partial Q}\right| _{r=r_{+}}=\frac{2 Q}{r_{+}^{2}},\left.
\frac{\partial f}{\partial r_{+}}\right| _{r=r_{+}}=4 \pi T,\left.
\frac{\partial f}{\partial P}\right| _{r=r_{+}}=\frac{8 \pi r_{+}^{2}}%
{3},\left. \frac{\partial f}{\partial a}\right| _{r=r_{+}}=-\frac{1}{r_{+}^{3
\omega_{q}+1}}.\label{parfn}%
\end{equation}
Bring Eqs. (\ref{fn}), (\ref{fnd}) and (\ref{parfn}) to Eq. $\left(
\ref{fextend}\right)  $ leads to
\begin{equation}
dM=2 \pi r_{+} Tdr_{+} +\frac{Q}{r_{+}}dQ +\frac{4}{3} \pi r_{+}^{3} dP
-\frac{1}{2r_{+}^{3 \omega_{q}}} da.\label{edM}%
\end{equation}
Using the Eqs. $\left(  \ref{varphi}\right)  $, $\left(  \ref{e11}\right)  $
and $\left(  \ref{e9}\right)  $, gives the first law of thermodynamics
\begin{equation}
dM=TdS+\varphi dQ+VdP+\eta da,\label{e12}%
\end{equation}
where $\eta$ is the physical quantity conjugate to the parameter $a$
\begin{equation}
\eta=-\frac{1}{2r_{+}^{3\omega_{q}}}.\label{e13}%
\end{equation}
And the Smarr relation \cite{Villalba:2019sxl} can be written as
\begin{equation}
M=2TS+\varphi Q-2VP+(1+3\omega_{q})\eta a.\label{e14}%
\end{equation}

For simplicity we use AdS radius $l$ instead of pressure $P$ do the next
calculation. Inserting Eqs. (\ref{e6}) and (\ref{e9}) into Eq. (\ref{a})
yields
\begin{equation}
dM=d(U+PV)=\omega(\omega-q\varphi)r_{+}^{2}dt+\frac{3r_{+}^{2}}{2l^{2}}
dr_{+}-\frac{r_{+}^{3}}{l^{3}}dl.\label{extendEQ}%
\end{equation}
Bring these results into the Eq. (\ref{edM}), we can obtain the variation of
the radius in horizon
\begin{equation}
d r_{+}=\frac{2 r_+ l^{2}(\omega-q \varphi)^{2}}{4 \pi l^{2} T-3 r_{+}} d t+\frac{l^{2} r_{+}^{-3 \omega_{q}-1}}{4 \pi l^{2} T-3 r_{+}} d a.
\end{equation}\label{e334}%
The variation of the entropy then takes on the form
\begin{equation}
d S=2 \pi r_{+} d r_{+}=\frac{4 \pi r_{+}^2 l^{2}(\omega-q \varphi)^{2}}{4 \pi l^{2} T-3 r_{+}} d t+\frac{2 \pi l^{2} r_{+}^{-3 \omega_{q}}}{4 \pi l^{2} T-3 r_{+}} d a.
\end{equation} 
It is not easy to determine the variation of entropy whether increase or
decrease. We can consider the extremal black hole which satisfies the
condition $T=0$. Suppose in the restricted extended phase space with $da > 0$, the change of the black hole entropy becomes
\begin{equation}
d S=2 \pi r_{+} d r_{+}=-\frac{4}{3} \pi r_+  l^{2}(\omega-q \varphi)^{2} d t-\frac{2 \pi}{3} l^{2} r_{+}^{-3 \omega_{q}-1} d a<0, \label{aa}
\end{equation}
which shows that the entropy decreases with time. So the second law of black
hole thermodynamics is not always valid for this situation.

Then we test the validity of the weak cosmic censorship conjecture in the
extremal and near-extremal RN-AdS black hole surrounded by quintessence. When
the charge absorbed by the black hole is more enough, the black hole is
overcharged and the weak cosmic censorship conjecture is violated. Therefore,
we just need check the existence of the event horizon after the scattering. A
simple method to check this existence is to evaluate the solution of the
equation $f(r)=0$. If the solution exists, the metric function $f(r)$ with a
minimum negative value guarantees the existence of the event horizon.

We suppose that there exists one minimum point at $r=r_{0}$ for $f(r)$, and
the minimum value of $f(r)$ is not greater than zero,
\begin{equation}
\delta\equiv f\left(  r_{0}\right)  \leq0,
\end{equation}
where $\delta=0$ corresponds to the extremal black hole. After the black hole
scatters the scalar field, the minimum point would move to $r_{0}+dr_{0} $,
the other parameters of the black hole change from $(M,Q,l,a)$ to
$(M+dM,Q+dQ,l+dl,a+da)$. The minimum value of $f(r)$ at $r=r_{0}+dr_{0}$ of
the final state is given by
\begin{align}
&  f\left(  r_{0}+dr_{0},M+dM,Q+dQ,l+dl,a+da\right) \nonumber\\
&  =\delta+\left.  \frac{\partial f}{\partial r_0}\right\vert _{r=r_{0}
}dr_0+\left.  \frac{\partial f}{\partial M}\right\vert _{r=r_{0}
}dM+\left.  \frac{\partial f}{\partial Q}\right\vert _{r=r_{0}}dQ+\left.
\frac{\partial f}{\partial l}\right\vert _{r=r_{0}}dl+\left.  \frac{\partial
f}{\partial a}\right\vert _{r=r_{0}}da\label{houwen}\\
&  =\delta+\delta_{1}+\delta_{2},\nonumber
\end{align}
where
\begin{equation}
\left.  \frac{\partial f}{\partial M}\right\vert _{r=r_{0}}=-\frac{2}{r_{0}},\left.  \frac{\partial f}{\partial Q}\right\vert _{r=r_{0}}=\frac{2Q}{r_{0}^{2}},\left.  \frac{\partial f}{\partial r_{0}}\right\vert_{r=r_{0}}=0,\left.  \frac{\partial f}{\partial l}\right\vert _{r=r_{0}}=-\frac{2r_{0}^{2}}{l^{3}},\left. \frac{\partial f}{\partial a}\right| _{r=r_{0}}=-\frac{1}{r_{0}^{3 \omega_{q}+1}}.\label{fds}
\end{equation}
Bring Eps. (\ref{fds}) and (\ref{edM}) to Eq. (\ref{houwen}) leads to
\begin{align}
\delta &  =f\left(  r_{0},M,Q,l,a\right)  ,\nonumber\\
\delta_{1}  &  =-\frac{2T}{r_{0}} dS+ \frac{2\left(  r_{+}^{3}-r_{0}%
^{3}\right)  }{l^{3}r_{0} }dl + \frac{\left(  r_{+}^{-3\omega_{q}}%
-r_{0}^{-3\omega_{q} }\right)  }{r_{0}} da,\\
\delta_{2}  &  =-\frac{2qQ\left(  r_{+}-r_{0}\right)  \left(  qQ-r_{+}
\omega\right)  }{r_{0}^{2}} dt.\nonumber
\end{align}

When it is an extreme black hole, $r_{0}=r_{+}$ and $T=0$, we can obtain
$\delta=0$, $\delta_{1}=0$ and $\delta_{2}=0$. Hence Eq. $\left(
\ref{houwen}\right)  $ can be rewritten as
\begin{align}
f \left(  r_{0}+dr_{0},M+dM,Q+dQ,l+dl,a+da\right)  =0,
\end{align}
which shows that the scattering does not change the minimum value. This
implies that the final state of the extremal black hole is still an extremal
black hole, the weak cosmic censorship conjecture is valid in the extremal
charged RN-AdS black hole surrounded by quintessence in the extended phase space.

When it is a near-extremal black hole, the order of the variables becomes
important, $f^{\prime}(r_{+})$ is very close to zero. To evaluate the value of
the above equation, we can let $r_{+}=r_{0}+\epsilon$, where $0<\epsilon\ll1$
and the relation $f(r_{+})=0$ and $f^{\prime}(r_{0})=0$. And then, the Eq.
$\left(  \ref{houwen}\right)  $ can be written as
\begin{align}
&  \delta<0,\nonumber\\
&  \delta_{1}=-\frac{f^{\prime\prime}\left(  r_{+}\right)  }{4\left(  \pi
r_{+}\right) } \epsilon dS +\frac{6r_{+}}{l^{3}} \epsilon dl -3\omega_{q}%
r_{+}^{-3\omega_{q}-2}\epsilon da +O\left(  \epsilon^{2}\right)  ,\nonumber\\
&  \delta_{2}=\frac{2qQ \left( \omega-q \varphi\right)  }{r_{+}} \epsilon dt
+O\left(  \epsilon^{2}\right) ,\label{benrong}%
\end{align}

where $dt$ is an infinitesimal scale and is set as $dt\sim\epsilon$. If the
initial black hole is near extremal, we have $dS\sim\epsilon$, $dl\sim
\epsilon$ and $da\sim\epsilon$. So $\delta_{1}+\delta_{2} \ll\delta$, the
final black hole has
\begin{equation}
f\left(  r_{0}+dr_{0},M+dM,Q+dQ,l+dl,a+da\right)  \approx\delta<0.\label{e44}%
\end{equation}
This indicates that the event horizon exists in the finial state. The black
hole can not be overcharged by the scattering of the scalar filed. Therefore,
the weak cosmic censorship conjecture is valid in the near-extremal charged
RN-AdS black hole surrounded by quintessence in the extended phase space.

\subsection{Normal phase space}

In the normal phase space, the cosmological constant and dimensional
parameters are fixed. The initial state of the black hole is represented by
$(M,Q,r_{+})$, and the final state is represented by $(M+dM,Q+dQ,r_{+}
+dr_{+})$. The variation of the radius can be obtained from the variation of
the metric function $f\left(  r_{+}\right)  $. For the initial state
$(M,Q,r_{+}) $, satisfies
\begin{equation}
f(M,Q,r_{+})=0,\label{f2}%
\end{equation}
We assume that the final state of the charged RN-AdS black hole is still a
black hole surrounded by quintessence, which satisfies
\begin{equation}
f(M+dM,Q+dQ,r_{+}+dr_{+})=0,\label{f1}%
\end{equation}
The functions $f(M+dM,Q+dQ,r_{+}+dr_{+})$ and $f(M,Q,r_{+})$ satisfy the
following relation
\begin{align}
&  f\left( M+d M, Q+d Q, r_{+}+d r_{+}\right) \nonumber\\
= &  f\left( M, Q, r_{+}\right) +\left. \frac{\partial f}{\partial M}\right|
_{r=r_{+}} d M+\left. \frac{\partial f}{\partial Q}\right| _{r=r_{+}} d
Q+\left. \frac{\partial f}{\partial r_{+}}\right| _{r=r_{+}} d r_{+}%
,\label{e32}%
\end{align}
Bring Eqs. (\ref{f1}), (\ref{f2}), (\ref{parfn}) into Eq. $\left(
\ref{e32}\right)  $ leads to
\begin{equation}
dM=\frac{Q}{r_{+}}dQ+2\pi Tr_{+}dr_{+}.\label{eem}%
\end{equation}
Using the Eqs. $\left(  \ref{varphi}\right)  $ and $\left(  \ref{e11}\right)
$ , gives the first law of thermodynamics%

\begin{equation}
dM=TdS+\varphi dQ.\label{e36}%
\end{equation}

In the normal phase space, Eq. $\left(  \ref{e30}\right)  $ gives the
transferred energy and charge within the certain time interval
\begin{equation}
dM=\omega(\omega-q\varphi)r_{+}^{2}dt,dQ=q(\omega-q\varphi)r_{+}%
^{2}dt.\label{normalEQ}%
\end{equation}
Bring these results into the Eq. $\left(  \ref{e32}\right)  $, we can obtain
the variation of the radius at horizon
\begin{equation}
dr_{+}=\frac{r_{+}}{2\pi T}(\omega-q\varphi)^{2}dt.\label{e34}%
\end{equation}
So the variation of the entropy takes on the form
\begin{equation}
dS=2\pi r_{+}dr_{+}=\frac{r_{+}^{2}}{T}(\omega-q\varphi)^{2}dt>0,\label{e35}%
\end{equation}
which shows that the entropy of the black hole increases. So the second law of
black hole thermodynamics is satisfied for the black hole.

Then we test the validity of the weak cosmic censorship conjecture in the
extremal and near-extremal via the black hole in this case. We suppose that
there exists one minimum point at $r=r_{0}$ for $f(r)$, and the minimum value
of $f(r)$ is not greater than zero
\begin{equation}
\delta\equiv f\left(  r_{0}\right)  \leq0,
\end{equation}
where $\delta=0$ corresponds to the extremal black hole. After the black hole
scatters the scalar field, the minimum point would move to $r_{0}+dr_{0} $.
For the final black hole solution, if the minimum value of $f(r)$ at
$r=r_{0}+dr_{0}$ is still not greater than zero, there exists an event
horizon. Otherwise, the final black hole solution is over the extremal limit,
and the weak cosmic censorship conjecture is violated.  For the final state, the minimum value of $f(r)$ at $r=r_{0}+dr_{0}$ becomes
\begin{align}
&  f\left(  r_{0}+dr_{0},M+dM,Q+dQ\right) \nonumber\\
&  =\delta+\left.  \frac{\partial f}{\partial r}\right\vert _{r=r_{0}}
dr_{0}+\left.  \frac{\partial f}{\partial M}\right\vert _{r=r_{0}}dM+\left.
\frac{\partial f}{\partial Q}\right\vert _{r=r_{0}}dQ.\label{fsd}%
\end{align}
Bring Eqs. (\ref{normalEQ}) and (\ref{fds}) into Eq. (\ref{fsd}), we can
obtain
\begin{equation}
f\left(  r_{0}+dr_{0},M+dM,Q+dQ\right)  =\delta-\frac{2}{r_{0}}\left(
\omega-\frac{qQ}{r_{+}}\right)  \left(  \omega-\frac{qQ}{r_{0}}\right)
r_{+}^{2}dt.\label{efrnormal}%
\end{equation}

When it is an extreme black hole, $r_{0}=r_{+}$ and $\delta=0$. So the above
equation becomes
\begin{equation}
f\left(  r_{0}+dr_{0}, M+dM,Q+dQ \right)  =-2\left(  \omega- q\varphi\right)
^{2}r_{+}dt<0,
\end{equation}
which indicates that the horizon exists in the finial state. The black hole
can not be overcharged by the scattering of the scalar filed. The extremal
black hole still extremal black hole if $\omega=q\varphi$, but the extremal
black hole becomes non-extremal black hole if $\omega\ne q\varphi$.

When it is a near-extremal black hole, the Eq. (\ref{efrnormal}) above
equation can be regarded as a quadratic function of $\omega$. When
$\omega=2qQ(\frac{1}{r_{+}}+\frac{1}{r_{0}})$, we find an maximum value on the
function of $f\left(  r_{0}+dr_{0},M+dM,Q+dQ\right) $
\begin{equation}
f\left(  r_{0}+dr_{0},M+dM,Q+dQ\right) _{max} =\delta-\frac{dtq^{2}%
Q^{2}\left(  r_{+}-r_{0}\right)  {}^{2}}{2r_{0}^{3}}<0.
\end{equation}
This implies metric function has a minimum negative value, which indicates
that the event horizon also exists in the finial state. The black hole can not
be overcharged by the scattering of the scalar filed. Therefore, the weak
cosmic censorship conjecture is valid in the near-extremal charged RN-AdS
black hole surrounded by quintessence in this case.

\section{Conclusion}

In this paper, we first derived the RN-AdS black hole surrounded by
quintessence via the scattering of a complex scalar field. The variations of
this black hole's energy and charge within the certain time interval can be
calculated. Then we investigated the validity of the thermodynamic laws in the
extended and normal phase spaces with these variations. The first law of
thermodynamics is always satisfied. The second law of thermodynamics is
satisfied in the normal phase space, but is not valid in the extended phase space.

Moreover, we test the validity of the weak cosmic censorship conjecture in the
extremal and near-extremal RN-AdS black hole surrounded by quintessence. The
weak cosmic censorship conjecture is valid both in the extended and normal
phase spaces. In the extended phase space, the extremal black hole stays
extremal under the scattering of the field. But in the normal phase space, the
extremal black hole becomes non-extremal black hole if $\omega\neq q\varphi$,
and the extremal black hole stays extremal if $\omega=q\varphi$.

\begin{acknowledgments}
We are grateful to Deyou Chen, Haitang Yang and Peng Wang for useful
discussions. This work is supported in part by NSFC (Grant No. 11375121,
11747171, 11747302 and 11847305). Natural Science Foundation of Chengdu
University of TCM (Grants No. ZRYY1729 and ZRQN1656). Discipline Talent
Promotion Program of /Xinglin Scholars(Grant No. QNXZ2018050) and the key fund
project for Education Department of Sichuan (Grant No. 18ZA0173). Sichuan
University students platform for innovation and entrepreneurship training
program (Grant No. C2019104639).
\end{acknowledgments}

\end{document}